\documentclass[12pt]{article}

\usepackage{amsmath}
\usepackage{amssymb}
\usepackage{amsfonts}
\usepackage{latexsym}
\usepackage{color}

\catcode `\@=11 \@addtoreset{equation}{section}

\catcode `\@=12

\newcommand{\be}{\begin{equation}}
\newcommand{\en}{\end{equation}}
\newcommand{\bea}{\begin{eqnarray}}
\newcommand{\ena}{\end{eqnarray}}
\newcommand{\beano}{\begin{eqnarray*}}
\newcommand{\enano}{\end{eqnarray*}}
\newcommand{\bee}{\begin{enumerate}}
\newcommand{\ene}{\end{enumerate}}

\newcommand{\N}{\mathfrak N}

\newcommand{\mc}{\mathcal}

\newcommand{\D}{{\mc D}}

\newcommand{\F}{{\cal F}}

\newcommand{\Lc}{{\cal L}}

\newcommand{\1}{1 \!\! 1}

\newcommand{\Hil}{\mc H}

\catcode `\@=11 \@addtoreset{equation}{section}
\catcode `\@=12

\begin{document}

\thispagestyle{empty}

\vspace*{2cm}

\begin{center}
{\Large \bf  Two-parameters pseudo-bosons}\\[10mm]

{\large F. Bagarello}\\
  Dipartimento di Metodi e Modelli Matematici,
Facolt\`a di Ingegneria,\\ Universit\`a di Palermo, I-90128  Palermo, Italy\\
e-mail: bagarell@unipa.it\\ Home page:
www.unipa.it/$^\sim$bagarell\\

\vspace{3mm}

\end{center}

\vspace*{2cm}

\begin{abstract}
\noindent We construct a two-parameters example of
{\em pseudo-bosons}, and we show that they are not regular, in the sense previously introduced by the author. In particular, we show that two biorthogonal bases of $\Lc^2(\Bbb R)$ can be constructed, which are not Riesz bases, in general.

\end{abstract}

\vspace{2cm}

%{\bf PACS Numbers}:  .......

\vfill

%\pagenumbering{roman}

\newpage

\section{Introduction}

In a series of recent papers \cite{bagpb1}-\cite{bagrev}, we have investigated some mathematical and physical aspects of the
so-called {\em pseudo-bosons} (PB),  originally introduced by Trifonov
in \cite{tri}. They arise from the canonical commutation relation (CCR)
$[a,a^\dagger]=\1$ upon replacing $a^\dagger$ by another (unbounded)
operator $b$ not (in general) related to $a$: $[a,b]=\1$. We have
shown that, under suitable assumptions, $N:=ba$ and $\N:=N^\dagger=a^\dagger b^\dagger$ can be both
diagonalized, and that their spectra coincide with the set of
natural numbers (including 0), ${\Bbb N}_0$. However the sets of
related eigenvectors are not orthonormal (o.n.) bases but they are automatically {\em biorthogonal}. In most of the
examples considered so far, they are bases of the Hilbert space of the system,
$\Hil$, and, in some cases, they turn out to be {\em Riesz bases}.

In this paper we discuss a two-parameters example of pseudo-bosons, showing, in particular, that they are not {\em regular} in the sense of \cite{bagpb4}. The paper is organized as follows: in the rest of this section we
briefly review $1$-dimensional PB.  In Sections II we discuss our example, while Section III contains our conclusions.

\subsection{A brief resume}

Let $\Hil$ be a given Hilbert space with scalar product $\left<.,.\right>$ and related norm $\|.\|$. Let $a$ and $b$  be two operators acting on $\Hil$ and satisfying the  commutation rule
\be
[a,b]=\1.
\label{21}
\en
Of course, this reduces to the CCR if $b=a^\dagger$. It is well known that $a$ and $b$ cannot be both bounded operators, so that they cannot be defined in all of $\Hil$. In the rest of the paper, given a certain operator $X$, we will call $D(X)$ its domain and $D^\infty(X):=\cap_{k\geq0}D(X^k)$ the domain of all its powers. In \cite{bagpb1}-\cite{bagrev} we have considered the following working assumptions:

\vspace{2mm}

{\bf Assumption 1.--} there exists a non-zero $\varphi_0\in\Hil$ such that $a\varphi_0=0$ and $\varphi_0\in D^\infty(b).$

\vspace{2mm}

{\bf Assumption 2.--} there exists a non-zero $\Psi_0\in\Hil$ such that $b^\dagger\Psi_0=0$ and $\Psi_0\in D^\infty(a^\dagger).$

\vspace{2mm}\noindent

If these hold we can introduce the vectors
\be
\varphi_n=\frac{1}{\sqrt{n!}}\,b^n\,\varphi_0, \quad \mbox{ and }\quad \Psi_n=\frac{1}{\sqrt{n!}}(a^\dagger)^n\Psi_0,
\label{22}\en
which clearly belong to $\Hil$ for all $n\geq 0$, and the  unbounded operators $N:=ba$ and $\N:=N^\dagger=a^\dagger b^\dagger$. In \cite{bagpb1} we have seen that $\varphi_n\in D(N)$, $\Psi_n\in D(\N)$, and
\be
N\varphi_n=n\varphi_n, \qquad \N\Psi_n=n\Psi_n,\qquad n\geq0.
\label{23}\en

In the above assumptions we have $\left<\Psi_n,\varphi_m\right>=\delta_{n,m}\left<\Psi_0,\varphi_0\right>$, for all $n, m\geq0$, which, if we fix the normalizations in such a way that  $\left<\Psi_0,\varphi_0\right>=1$, becomes
\be
\left<\Psi_n,\varphi_m\right>=\delta_{n,m}, \quad \forall n,m\geq0.
\label{27}\en
This means that the $\Psi_n$'s and the $\varphi_n$'s are biorthonormal.

Calling $\D_\varphi$ and $\D_\Psi$ respectively the linear span of  $\F_\varphi=\{\varphi_n,\,n\geq 0\}$ and $\F_\Psi=\{\Psi_n,\,n\geq 0\}$, and $\Hil_\varphi$ and $\Hil_\Psi$ their closures, we can  prove  that $\Hil_\varphi\subseteq\Hil$ and $\Hil_\Psi\subseteq\Hil$. However, in most examples reviewed in \cite{bagrev}, these three Hilbert spaces  coincide so that it is natural to require the following

\vspace{2mm}

{\bf Assumption 3.--} $\Hil_\varphi=\Hil_\Psi=\Hil$.

\vspace{2mm}

Hence, both $\F_\varphi$ and $\F_\Psi$ are bases in $\Hil$. The resolution of the identity looks now
\be
\sum_{n=0}^\infty
|\varphi_n><\Psi_n|=\sum_{n=0}^\infty
|\Psi_n><\varphi_n|=\1,
\label{211}\en
where $\1$ is the  identity of $\Hil$ and where the useful Dirac bra-ket notation has been adopted.  Let us
now introduce the operators $S_\varphi$ and $S_\Psi$ via their
action respectively on  $\F_\Psi$ and $\F_\varphi$: \be
S_\varphi\Psi_{ n}=\varphi_{ n},\qquad
S_\Psi\varphi_{ n}=\Psi_{ n}, \label{213}\en for all $ n$, which in particular imply that
$\Psi_{ n}=(S_\Psi\,S_\varphi)\Psi_{ n}$ and
$\varphi_{ n}=(S_\varphi \,S_\Psi)\varphi_{ n}$, for all
$ n$. Hence \be S_\Psi\,S_\varphi=S_\varphi\,S_\Psi=\1 \quad
\Rightarrow \quad S_\Psi=S_\varphi^{-1}. \label{214}\en In other
words, both $S_\Psi$ and $S_\varphi$ are invertible and one is the
inverse of the other. Furthermore, we can also check that they are
both positive, well defined and symmetric, \cite{bagpb1}. Moreover, at
least formally, it is possible to write these operators in the
bra-ket notation as \be S_\varphi=\sum_{ n}\,
|\varphi_{ n}><\varphi_{ n}|,\qquad S_\Psi=\sum_{ n}
\,|\Psi_{ n}><\Psi_{ n}|. \label{212}\en
 These expressions are
only formal, at this stage, since the series may or may not converge in
the uniform topology and the operators $S_\varphi$ and $S_\Psi$ could be unbounded.
Indeed we know,  \cite{you}, that two biorthogonal bases are related by a bounded operator, with bounded inverse, if and only if they are Riesz bases\footnote{Recall that a set of vectors $\phi_1, \phi_2 , \phi_3 , \; \ldots \; ,$ is a Riesz basis of a Hilbert space $\mathcal H$, if there exists a bounded operator $V$, with bounded inverse, on $\mathcal H$, and an orthonormal basis of $\Hil$,  $\varphi_1, \varphi_2 , \varphi_3 , \; \ldots \; ,$ such that $\phi_j=V\varphi_j$, for all $j=1, 2, 3,\ldots$}. This is why in \cite{bagpb1} we have also considered

\vspace{2mm}

{\bf Assumption 4.--} $\F_\varphi$ and $\F_\Psi$ are  both Riesz bases.

\vspace{2mm}

This implies  that $S_\varphi$ and $S_\Psi$ are bounded operators, so that their domains can be taken to be all of $\Hil$. Particles satisfying Assumptions 1, 2 and 3 are called {\em pseudo-bosons} (PB), while if they also satisfy Assumption 4, they are called {\em regular pseudo-bosons} (RPB), \cite{bagpb4}.

We end this short review recalling that the following intertwining relations are satisfied:
 \be S_\Psi\,N=\N S_\Psi \quad \mbox{ and }\quad
N\,S_\varphi=S_\varphi\,\N. \label{219}\en  This is
related to the fact that the spectra of $N$ and $\N$
coincide and that their eigenvectors are related by the operators
$S_\varphi$ and $S_\Psi$, see equations (\ref{23}) and (\ref{213}), in agreement with the literature on
intertwining operators, \cite{intop,bag1}.

\section{The model}

Let $\epsilon$ and $\eta$ be respectively a real and a complex parameter and let us consider the following self-adjoint, unbounded and invertible operator $S_{\epsilon,\eta}:=\exp\{\epsilon\,a^\dagger\,a+\eta\,a^2+\overline{\eta}\,{a^\dagger}^2\}$.  We will assume in the following that $\epsilon^2>4|\eta|^2$.

\vspace{2mm}
{\bf Remark:--} $S_{\epsilon,\eta}$ extends formally the operator $T_\theta$ introduced in \cite{bagpb4}, where $\epsilon=0$ and $\eta$ was purely imaginary. However, under these conditions, $\epsilon^2>4|\eta|^2$ is never satisfied.
\vspace{2mm}

 We can define two new operators as follows:
\be
\left\{
\begin{array}{ll}
A_{\epsilon,\eta}:=S_{\epsilon,\eta}\,a\,S_{\epsilon,\eta}^{-1}=\left(\cosh\theta-\frac{\epsilon}{\theta}\,
\sinh\theta\right)a-2\,\frac{\overline{\eta}}{\theta}\,
\sinh\theta\,a^\dagger,\\
B_{\epsilon,\eta}:=S_{\epsilon,\eta}\,a^\dagger\,S_{\epsilon,\eta}^{-1}=2\,\frac{{\eta}}{\theta}\,
\sinh\theta\,a+\left(\cosh\theta\frac{\epsilon}{\theta}\,
\sinh\theta\right)a^\dagger,
\end{array}
\right.
\label{31}\en
where $\theta=\sqrt{\epsilon^2-4|\eta|^2}$ is real and positive and $[a,a^\dagger]=\1$. It is clear that, in general, $B_{\epsilon,\eta}^\dagger\neq A_{\epsilon,\eta}$. It is also clear that $[A_{\epsilon,\eta},B_{\epsilon,\eta}]=\1$. Hence we might have to do with PB or even with RPB. In the rest of this section we will show that, at least for certain values of $\epsilon$ and $\eta$, we have indeed PB, but we never get RPB (except when $\epsilon=\eta=0$, choice which is not compatible with our previous assumption and which would return ordinary bosons). To simplify the notation we will omit, from now on, the suffixes $\epsilon$ and $\eta$. To prove that Assumptions 1, 2 and 3 are satisfied it is convenient to work in the coordinate representation, so that $\Hil=\Lc^2(\Bbb R)$. Then, since $a=\frac{1}{\sqrt{2}}\left(x+\frac{d}{dx}\right)$ and $a^\dagger=\frac{1}{\sqrt{2}}\left(x-\frac{d}{dx}\right)$, we can rewrite $A$ and $B$ as follows:
\be
\left\{
\begin{array}{ll}
A=k_{A,+}\,x+k_{A,-}\,\frac{d}{dx}, \qquad B=k_{B,+}\,x+k_{B,-}\,\frac{d}{dx}\\
k_{A,\pm}=\frac{1}{\sqrt{2}}\left(k_{A,1}\pm \,k_{A_2}\right), \qquad k_{B,\pm}=\frac{1}{\sqrt{2}}\left(k_{B,1}\pm \,k_{B_2}\right),\\
k_{A,1}=\cosh\theta-\frac{\epsilon}{\theta}\,
\sinh\theta,\quad k_{A_2}=-2\,\frac{\overline{\eta}}{\theta}\,
\sinh\theta,\\
k_{B,1}=2\,\frac{{\eta}}{\theta}\,
\sinh\theta, \qquad k_{B,2}=\cosh\theta+\frac{\epsilon}{\theta}\,
\sinh\theta.
\end{array}
\right.
\label{32}\en
The solution of equation $A\varphi_0(x)=0$ is easily found: $\varphi_0(x)=N_0^\varphi\,\exp\{-\frac{1}{2}\,
\frac{k_{A,+}}{k_{A,-}}\,x^2\}$, $N_0^\varphi$ being a normalization constant to be fixed. Analogously, the solution of $B^\dagger\Psi_0(x)=0$ is $\Psi_0(x)=N_0^\Psi\,\exp\{\frac{1}{2}\,\overline{\left(\frac{k_{B,+}}{k_{B,-}}\right)}\,x^2\}$, and $N_0^\Psi$ is (partially) fixed by requiring that $\left<\varphi_0,\Psi_0\right>=1$. Of course, this condition is meaningful if, for instance, $\varphi_0(x)$ and $\Psi_0(x)$ are both square-integrable. This is ensured if the following conditions are satisfied:
\be
\Re\left(\frac{k_{A,+}}{k_{A,-}}\right)>0,\qquad \Re\left(\frac{k_{B,+}}{k_{B,-}}\right)<0.
\label{33}\en
If these inequalities hold, a straightforward computation, and the  equality $k_{A,-}\,k_{B,+}-k_{A,+}k_{B,-}=1$, produces the following result:
\be
 \left\{
\begin{array}{ll}\varphi_n(x)=\frac{1}{\sqrt{n!\,}}\,B^n\,\varphi_0(x)=\frac{N_0^\varphi}{\sqrt{n!\,}}\,p_n^\varphi(x)\,\exp\{-\frac{1}{2}\,
\frac{k_{A,+}}{k_{A,-}}\,x^2\},\\
 \Psi_n(x)=\frac{1}{\sqrt{n!\,}}\,{A^\dagger}^n\Psi_0(x)=\frac{N_0^\Psi}{\sqrt{n!\,}}
 \,p_n^\Psi(x)\,\exp\{\frac{1}{2}\,\overline{
\left(\frac{k_{B,+}}{k_{B,-}}\right)}\,x^2\},
\end{array}
\right.
\label{34}\en
where  $p_n^\varphi(x)$ and $p_n^\Psi(x)$ are n-th order polynomials defined recursively as: $p_0^\varphi(x)=1$, $p_{n+1}^\varphi(x)=\frac{1}{k_{A,-}}\,x\,p_n^\varphi(x)+k_{B,-}\,\frac{d p_n^\varphi(x)}{dx}$ and $p_0^\Psi(x)=1$, $p_{n+1}^\Psi(x)=\frac{-1}{\overline{k_{B,-}}}\,x\,p_n^\Psi(x)-\overline{k_{A,-}}\,\frac{d p_n^\Psi(x)}{dx}$.

Notice that these formulas  prove that $\varphi_0(x)\in D^\infty(B)$ and $\Psi_0(x)\in D^\infty(A^\dagger)$, since polynomials times gaussians are surely functions of $\Lc^2(\Bbb{R})$. Hence, Assumptions 1 and 2 are satisfied. As for Assumption 3, the same arguments given in \cite{bagpb2} and \cite{bagpb4} apply. This allows us to conclude that our operators $A$ and $B$ give rise to PB. Now the question is the following: are these PB also regular? In other words, is Assumption 4 satisfied?

The (maybe) simplest way to answer this question makes use of the non self-adjoint operator $H=\omega\, BA$, where $\omega$ is a fixed real constant, which can be written as $H=\omega N$. Its adjoint is $H^\dagger=\omega\N$, and we know that $\varphi_n$ and $\Psi_n$ are their respective eigenstates: $H\,\varphi_n=\omega\,n\,\varphi_n$ and $H^\dagger\,\Psi_n=\omega\,n\,\Psi_n$, $n\geq0$. Let us now introduce a self-adjoint operator $h=\omega \,a^\dagger a$, whose eigenvectors are very well known: $\Phi_n=\frac{{a^\dagger}^n}{\sqrt{n!\,}}\,\Phi_0$, where $a\Phi_0=0$. We have $h\,\Phi_n=\omega\,n\,\Phi_n$, $n\geq0$.

The first remark is that the operator $S$ intertwines between $H$ and $h$, and between $H^\dagger$ and $h$: $HS=Sh$ and $SH^\dagger=hS$. This is in agrement, \cite{intop}, with the fact that the spectra of $H$, $H^\dagger$ and $h$ all coincide.

It is now easy to check that a single complex constant $\gamma$ exists such that
\be
\varphi_n=\gamma\,S\,\Phi_n,\qquad \Psi_n=\frac{1}{\gamma}\,S^{-1}\,\Phi_n,
\label{35}\en
for all $n\geq0$. Let us prove the first equality. Recalling the definition of $\varphi_n$ and the expression of $B$ in (\ref{31}) we find that $\varphi_n=\frac{1}{\sqrt{n!\,}}\,B^n\,\varphi_0=S\frac{1}{\sqrt{n!\,}}\,{a^\dagger}^n\,S^{-1}\varphi_0
$. Notice now that $a\,S^{-1}\varphi_0=S^{-1}\left(S\,a\,S^{-1}\right)\varphi_0=S^{-1}\,A\,\varphi_0=0$. Hence $S^{-1}\varphi_0$ must be proportional to $\Phi_0$ (assuming no degeneracy of the lowest energetic level of $h$): $S^{-1}\varphi_0=\gamma\Phi_0$, for some complex $\gamma$. Going back to our previous expression for $\varphi_n$, we deduce that  $\varphi_n=S\frac{1}{\sqrt{n!\,}}\,{a^\dagger}^n\,S^{-1}\varphi_0=\gamma S\frac{1}{\sqrt{n!\,}}\,{a^\dagger}^n\,\Phi_0
$, so that the first equality in (\ref{35}) follows. The second equality can be proved in similar way. An immediate consequence of (\ref{35}) is that $\F_\varphi$ and $\F_\Psi$ are obtained acting on the o.n. basis $\{\Phi_n,\,n\geq0\}$ with an unbounded operator. This implies, \cite{you}, that they are not Riesz bases. Hence we have constructed PB but not RPB.

\subsection{On conditions (\ref{33})}

The crucial assumption behind our results above is that the inequalities in (\ref{33}) are both satisfied. Otherwise the solutions of $A\varphi_0(x)=B^\dagger\Psi_0(x)=0$ will not be in $\Lc^2(\Bbb{R})$ and no biorthogonal sets of $\Hil$ could be constructed. We will now briefly show that these two inequalities admit common solutions. For that, and to simplify the treatment, we assume $\eta$ to be real and we write $\epsilon$ as $\epsilon=\alpha\,\eta$, for some real $\alpha$ with $\alpha^2>4$. This is because we want to satisfy our original assumption $\epsilon^2>4|\eta|^2$. Then minor computations show that $\Re\left(\frac{k_{A,+}}{k_{A,-}}\right)>0$ if and only if
$$
\frac{1-\sqrt{\frac{\alpha+2}{\alpha-2}}\,\tanh(\eta\sqrt{\alpha^2-4})}
{1-\sqrt{\frac{\alpha-2}{\alpha+2}}\,\tanh(\eta\sqrt{\alpha^2-4})}\,>0,
$$
while $\Re\left(\frac{k_{B,+}}{k_{B,-}}\right)<0$ if and only if
$$
\frac{1+\sqrt{\frac{\alpha+2}{\alpha-2}}\,\tanh(\eta\sqrt{\alpha^2-4})}
{1+\sqrt{\frac{\alpha-2}{\alpha+2}}\,\tanh(\eta\sqrt{\alpha^2-4})}\,>0.
$$
It is a simple exercise in calculus to check that a common solution does exist for $\eta\in]-\eta_0,\eta_0[$, where $\eta_0=\frac{1}{\sqrt{\alpha^2-4}}\,\tanh^{-1}\left(\sqrt{\frac{\alpha-2}{\alpha+2}}\right)$. Hence for each $\eta$ taken in this range and $\alpha$ such that $\alpha^2>4$, fixing $\epsilon=\alpha\,\eta$, Assumptions 1 and 2, and 3 as a consequence, are satisfied.

\vspace{3mm}

Even if condition $\epsilon^2>4\eta^2$ does not allow us to simply fix $\epsilon=\eta=0$, which, as already remarked, would give back ordinary bosons, we can nevertheless consider the limit $\epsilon\rightarrow0$, $\eta\rightarrow0$, in such a way the inequality above is satisfied. In this case we get $\varphi_n(x)=\frac{N_0^\varphi}{\sqrt{n!\,2^n}}\,H_n(x)\,\exp\{-\frac{1}{2}\,x^2\}$ and $\Psi_n(x)=\frac{N_0^\Psi}{\sqrt{n!\,2^n}}\,H_n(x)\,\exp\{-\frac{1}{2}\,x^2\}$, as expected. This is not the only case in which the Hermite polynomials $H_n(x)$ are recovered: indeed, if $k_{A,-}k_{B,-}=-\,\frac{1}{2}$, we obtain
$$
\varphi_n(x)=\frac{N_0^\varphi}{\sqrt{n!\,}}\,(-k_{B,-})^n\,H_n(x)\,\exp\left\{-\frac{1}{2}\,
\frac{k_{A,+}}{k_{A,-}}\,x^2\right\}
$$
and
$$
\Psi_n(x)=\frac{N_0^\Psi}{\sqrt{n!\,}}\,(\overline{k_{A,-}})^n\,H_n(x)\,\exp\left\{\frac{1}{2}\,\overline{
\left(\frac{k_{B,+}}{k_{B,-}}\right)}\,x^2\right\},
$$
$n\geq0$, which return the previous solutions if $k_{A,-}=-k_{B,-}=\frac{1}{\sqrt{2}}$ (which correspond to ordinary bosons). Incidentally we observe that condition $k_{A,-}k_{B,-}=-\,\frac{1}{2}$ admits solutions other than this. A direct computation shows that, for instance, $\eta=0$ is such a solution. Other solutions can be recovered by considering the limit $(\epsilon-2\eta)\rightarrow0$.

\section{Conclusions}

In this paper we have reviewed some basic facts about PB and RPB, giving also an example based on a two-parameters (one real and the second complex) deformation of the standard creation and annihilation operators. Interesting features appear, like the relation with Hermite polynomials, under suitable conditions. Other examples can be found in \cite{bagrev}, and more are still under construction.

\section*{Acknowledgements}
   The author wants to thank Miroslav Znojil for his kind invitation to write this short note. The author also acknowledge M.I.U.R. for financial support.

\end{document}